\documentclass[english,prl,notitlepage,reprint]{revtex4-1}
\usepackage[T1]{fontenc}
\usepackage[latin9]{inputenc}
\usepackage{amsmath}
\usepackage{amssymb}
\usepackage{graphicx}
\usepackage{babel}
\begin{document}

\title{Unraveling open-system quantum dynamics of non-interacting Fermions}

\author{Zhu Ruan}

\affiliation{Fritz Haber Center for Molecular Dynamics and Institute of Chemistry,
The Hebrew University of Jerusalem, Jerusalem 9190401, Israel}

\author{Roi Baer{*}}

\affiliation{Fritz Haber Center for Molecular Dynamics and Institute of Chemistry,
The Hebrew University of Jerusalem, Jerusalem 9190401, Israel}
\begin{abstract}
The Lindblad equation is commonly used for studying quantum dynamics
in open systems that cannot be completely isolated from an environment,
relevant to a broad variety of research fields, such as atomic physics,
materials science, quantum biology and quantum information and computing.
For electrons in condensed matter systems, the Lindblad dynamics is
intractable even if their mutual Coulomb repulsion could somehow be
switched off. This is because they would still be able to affect each
other by interacting with the bath. Here, we develop an approximate
approach, based on the Hubbard-Stratonovich transformation, which
allows to evolve non-interacting Fermions in open quantum systems.
We discuss several applications for systems of trapped 1D Fermions
showing promising results. 
\end{abstract}
\maketitle

\section{Introduction}

Decoherence, dephasing and dissipation in large open quantum systems
are important phenomena in a broad variety of fields, such as nonadiabatic
processes in chemistry and materials science, \cite{Head-Gordon1995,schaller2005breaking,baer2003therole,Baer2006,Gdor2015,shenvi2012nonadiabatic,dong2015observation},
quantum biology~\cite{collini2010coherently,romero2014quantum} and
quantum information~\cite{ingarden2013information,schlosshauer2007decoherence}.
They are commonly described using the concept of the density matrix
(DM), which generalizes the notion of a wave function as the quantum
state descriptor. Despite great success in atomic physics, DM approaches
have not found extensive application in the field of large electronic
systems, except in cases of small systems, where it is sufficient
and possible to address only a small number of electronic states \cite{prezhdo1999mean,muhlbacher2008real,abramavicius2009coherent,Esposito2009,wilner2013bistability,cohen2013numerically,schinabeck2016hierarchical}.
For describing the quantum dynamics of open systems having a large
number of electrons and electronic states a different approach is
probably needed. Here, it is natural to consider time-dependent (current)
density functional theory (TDDFT), based on the Runge-Gross (RG) theorem~\cite{Runge1984}
which simplifies the treatment of the dynamics of interacting electrons
by mapping them onto non-interacting Fermions. Extensions of the RG
theorem to open systems have indeed appeared \cite{Burke2005,Kurth2005,Zheng2007,Pershin2008,Yuen-Zhou2010},
but the follow-up progress has yet to be achieved, and the main cause
for delay is the fact that non-interacting Fermions develop an interaction
through the coupling with the bath.\footnote{This is true when the electrons interact with the bath through the
one-body density matrix, which the case of interest here and in most
practical applications. There exists an important class of problems
in which the electron interaction with the bath is ``linear'' with
the particle creation/destruction operators allowing an easier TDDFT
adaption (see \cite{Kurth2005}).}.

In this paper, we develop a method to describe the DM time evolution
of non-interacting Fermions (Section~\ref{sec:Fermion-Unraveling})
as they are coupled to an external bath. We work within the Lindblad
formalism~\cite{Lindblad1976,gorini1976completely,Breuer2002,schaller2014open},
which is useful for describing Markovian open system dynamics.. The
method makes use of the unraveling procedure, which transforms the
Lindblad equation on the DM into a random walk in wave functions space.
The effective Fermion-Fermion interactions induced by the bath are
converted into \emph{additional }random-walk terms. Applications of
the method, first to an analytically solvable model and then to a
system of trapped 1D Fermions in a double-well are given in Section~\ref{sec:Applications-to-trapped}.
We believe, that the present development forms a significant stepping
stone for applying TDDFT to the study of the dynamics of open electronic
systems in the future.

\begin{figure*}[t]
\begin{raggedright}
\includegraphics[width=1\textwidth]{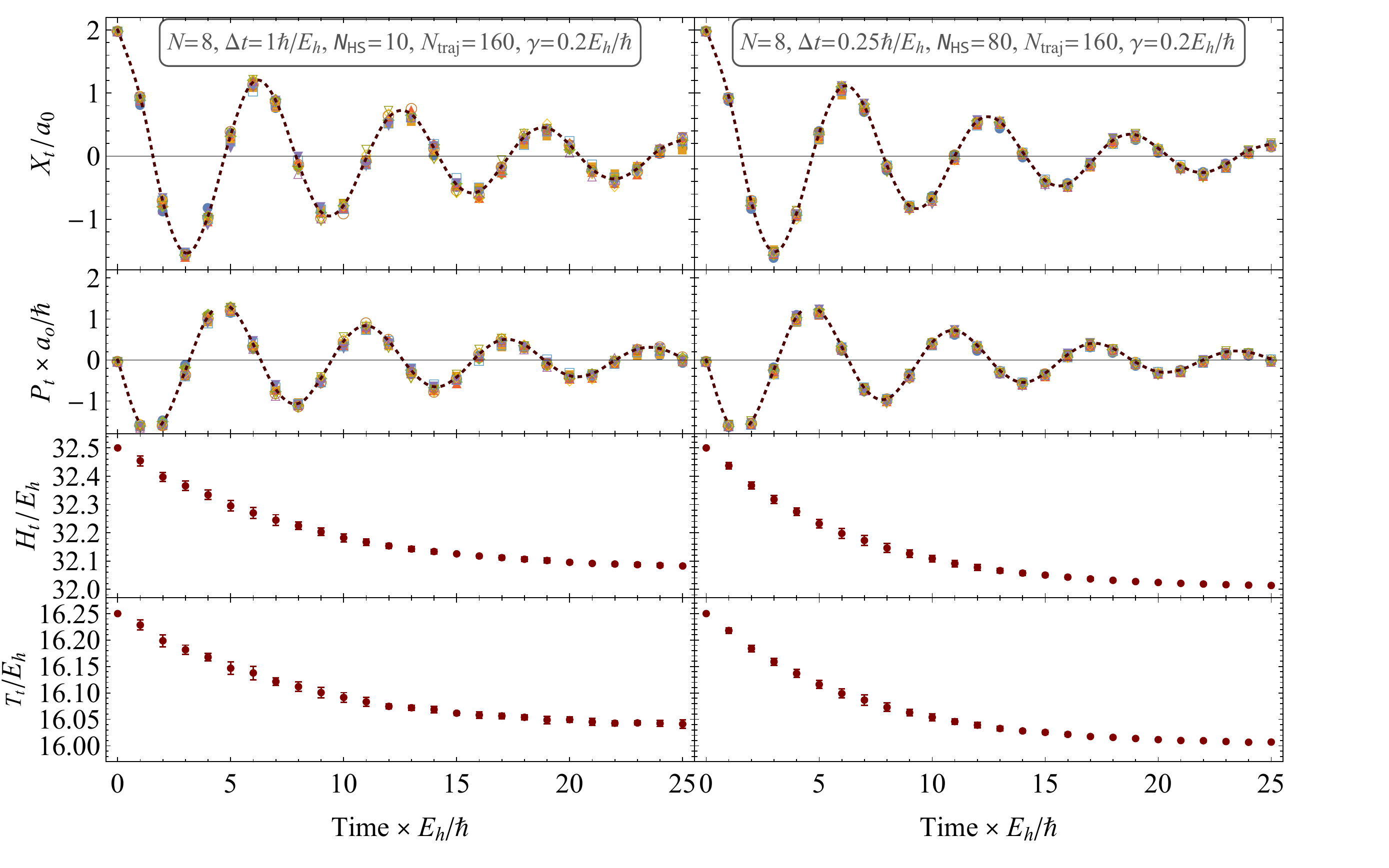}
\par\end{raggedright}
\caption{\label{fig:harmonic-XP}The time-dependent total displacement $X_{t}$,
total momentum $P_{t}$, total energy $H_{t}$ and total kinetic energy
$T_{t}$ transients for $8$ non-interacting Fermions in the Harmonic
trap of Section~\ref{subsec:A-harmonic-trap}, starting from the
pure state $\hat{\rho}_{\theta}=\left|\Phi_{\theta}\right\rangle \left\langle \Phi_{\theta}\right|$
with $\theta=\pi/4$. The $X_{t}$ and $P_{t}$ panels show analytical
transients (Eq.~(\ref{eq:xt})) as dashed lines while the calculated
results of 10 independent runs (each having 160 trajectories) are
shown as symbols. The $H_{t}$ and $T_{t}$ results are depicted as
statistical error bars centered on the average over the 10 runs. The
computed results shown in the left and right panels are based on different
time steps $\Delta t$ and number of HS iterations $N_{HS}$. }
\end{figure*}

\section{\label{sec:Fermion-Unraveling}Fermion Unraveling }

The density matrix (DM) operator $\hat{\rho}$ represents the quantum
state of a system, open or closed, generalizing the concept of a pure
wave function. It can be written in terms of its eigenvalues $w_{s}$
and eigenfunctions $\Phi^{s}$ as
\begin{equation}
\hat{\rho}=\sum_{s}p_{s}\left|\Phi^{s}\right\rangle \left\langle \Phi^{s}\right|
\end{equation}
and the eigenvalues $p_{s}$ being the probability for the system
to be in state $\Phi^{s}$. Clearly the DM must be Hermitean, positive-definite
($p_{s}>0$) and unit-traced ($\sum_{s}p_{s}=1$). If $\hat{O}$ is
an operator corresponding to an observable property, then the expectation
value of its measurement is expressed neatly as a trace: $O=tr\left[\hat{\rho}\hat{O}\right]=\sum_{s}p_{s}\left\langle \Phi_{s}\left|\hat{O}\right|\Phi_{s}\right\rangle $.
In any time-dependent process, the DM evolution is determined by an
equation of motion which must preserve its trace, its Hermiticity
and its positivity. The most general ``Markovian'' equation of motion
that respects these basic tenants is the so-called Lindblad-equation
\cite{Lindblad1976,Alicki2007,Breuer2002,schaller2014open}: 
\begin{equation}
\dot{\hat{\rho}}\left(t\right)=-\frac{i}{\hbar}\left[\hat{H},\hat{\rho}\right]+\text{\ensuremath{\mathfrak{D}}}\hat{\rho}\left(t\right)\label{eq:LindbladMasterEq}
\end{equation}
where $\hat{H}=\hat{H}^{\dagger}$ is the Hermitean \emph{effective
Hamiltonian }and the \emph{dissipative }part, which is of the form:
\begin{equation}
\text{\ensuremath{\mathfrak{D}}}\hat{\rho}=\left(\hat{L}^{\alpha}\hat{\rho}\hat{L}^{\alpha\dagger}-\frac{1}{2}\hat{L}^{\alpha\dagger}\hat{L}^{\alpha}\hat{\rho}-\frac{1}{2}\hat{\rho}\hat{L}^{\alpha\dagger}\hat{L}^{\alpha}\right)\label{eq:Disspitaor}
\end{equation}
where $\hat{L}^{\alpha}$ are \emph{Lindblad operators }($\alpha=1,2,\dots,N_{L}$
and we adopt the Einstein convention that repeated dummy indices get
summed). These equations of motions are supplemented by an initial
condition $w_{s}\left(0\right)$ and $\Phi_{0}^{s}$ at $t=0$. 

For many-body systems, working with the DM is difficult, if not impossible
and therefore unraveling procedures~\cite{carmichael1993quantum,wiseman1993quantum,dalibard1992wave,Gisin1992}
were developed where the DM is represented as an expected value involving
\emph{random }wave functions $\Phi\left(t\right)$ 
\begin{equation}
\hat{\rho}\left(t\right)=E\left\{ \frac{\left|\Phi\left(t\right)\right\rangle \left\langle \Phi\left(t\right)\right|}{\left\langle \Phi\left(t\right)\left|\Phi\left(t\right)\right.\right\rangle }\right\} .
\end{equation}
Each of the these states, $\Phi\left(t\right)$ start from a randomly
selected initial state $\Phi_{0}^{s}$ with probability $w_{s}$ and
is then evolved separately in time according to the following nonlinear
stochastic Schr�dinger equation:

\begin{align}
d\Phi\left(t\right) & =\left[-\frac{i}{\hbar}\hat{H}dt\right.\label{eq:unraveling-two-body}\\
 & \left.+\left(\left(\left\langle L^{\alpha}\right\rangle _{t}^{*}-\frac{1}{2}\hat{L}^{\alpha\dagger}\right)dt+dw_{\alpha}\right)\hat{L}^{\alpha}\right]\Phi\left(t\right)\nonumber 
\end{align}
where $dw_{\alpha}$ are independent Wiener processes with $E\left[dw^{\alpha}dw^{\beta}\right]=\delta^{\alpha\beta}dt$
and $\left\langle L^{\alpha}\right\rangle _{t}\equiv\frac{\left\langle \Phi\left(t\right)\left|\hat{L}^{\alpha}\right|\Phi\left(t\right)\right\rangle }{\left\langle \Phi\left(t\right)\left|\Phi\left(t\right)\right.\right\rangle }$~\cite{Gisin1992}
or $\left\langle L^{\alpha}\right\rangle _{t}=\frac{Re\left\langle \Phi\left(t\right)\left|\hat{L}_{\alpha}\right|\Phi\left(t\right)\right\rangle }{\left\langle \Phi\left(t\right)\left|\Phi\left(t\right)\right.\right\rangle }$.~\cite{wiseman1993quantum}
This equation is much easier to handle than Eq.~(\ref{eq:LindbladMasterEq})
since it involves only the wave function. But this comes with a sizable
price tag: a non-linear Schr�dinger equation combined with stochastic
noise.

The unraveling procedure given above applies to all Lindblad equations,
in particular for non-interacting Fermion systems, where the effective
Hamiltonian and the Lindblad operators are one-body operators: 
\begin{align}
\hat{H} & =\sum_{n}\hat{h}\left(n\right)\label{eq:H-hat}\\
\hat{L}^{\alpha} & =\sum_{n}\hat{\ell}^{\alpha}\left(n\right)\label{eq:L-alpha}
\end{align}
where 
\begin{equation}
\hat{h}=\frac{\hat{p}^{2}}{2m}+V\left(\hat{x}\right),\label{eq:single-particle-h}
\end{equation}
is a single electron Hamiltonian ($\hat{h\left(n\right)}$ is this
Hamiltonian applied for electron number $n$). One notices that the
term $\hat{L}^{\alpha\dagger}\hat{L}^{\alpha}=\sum_{nm}\hat{\ell}^{\alpha\dagger}\left(n\right)\hat{\ell}^{\alpha}\left(m\right)$
appearing in Eq.~\ref{eq:unraveling-two-body} is a two-body operator
and thus the unraveling of non-interacting electrons is essentially
an interacting electron problem. 

We make progress here through the Hubbard-Stratonovich transformation
\cite{stratonovich1957rl,Hubbard1959}, $e^{-\frac{1}{2}\hat{R}^{2}dt}\propto\int_{-\infty}^{\infty}e^{-\frac{\eta^{2}}{2dt}}e^{i\hat{R}\eta}d\eta$
which converts Eq.~(\ref{eq:unraveling-one-body}) into a new equation
involving a 3-component Brownian (Wiener) motion:
\begin{align}
d\Phi\left(t\right) & =-\frac{i}{\hbar}\left(\hat{H}dt-i\left(d\hat{H}_{R}+d\hat{H}_{C}+d\hat{H}_{S}\right)\right)\Phi\left(t\right)\label{eq:unraveling-one-body}
\end{align}
where:
\begin{align}
d\hat{H}_{R} & \equiv-\hbar\left[\left\langle L^{\alpha}\right\rangle _{t}^{*}dt+dw_{\alpha}+idu_{\alpha}\right]\hat{R}^{\alpha}\\
d\hat{H}_{S} & \equiv-\hbar\left[\left\langle L^{\alpha}\right\rangle _{t}^{*}dt+dw_{\alpha}+dv_{\alpha}\right]i\hat{S}^{\alpha}\\
d\hat{H}_{C} & \equiv\hbar\sum_{\alpha}\hat{C}^{\alpha}dt
\end{align}
where $\hat{R}^{\alpha}=Re\left[\hat{L}^{\alpha}\right]$ and $\hat{S}^{\alpha}=Im\left[\hat{L}^{\alpha}\right]$
, $\hat{C}^{\alpha}=i\left[\hat{R}^{\alpha},\hat{S}^{\alpha}\right]_{-}$
are three Hermitean one-particle operators (so that $\hat{L}^{\alpha\dagger}\hat{L}^{\alpha}=\hat{R}^{\alpha}\hat{R}^{\alpha}+\hat{S}^{\alpha}\hat{S}^{\alpha}+\hat{C}$)
and where like $dw_{\alpha}$, also $du_{\alpha}$, $dv_{\alpha}$
are each a Wiener processes i.e. a random number drawn from the normal
distribution with mean zero and variance $dt$. There is, however,
an important, delicate, point here: for each $dw_{\alpha}$ we must
sample $du_{\alpha}$ and $dv_{\alpha}$ many times so as to enable
an accurate calculation of $\left\langle L^{\alpha}\right\rangle _{t}$.
Hence, the algorithm we use to evolve the DM of non-interacting Fermions
is as follows:
\begin{enumerate}
\item Assume we have the Slater wave function $\Phi\left(t\right)=\det\left[\phi_{1}\left(t\right)\cdots\phi_{N}\left(t\right)\right]$
and the expected values $L^{\alpha}\left(t\right)$.
\item Propagate from $t\to t+dt$:
\begin{enumerate}
\item Sample $dw_{\alpha}$.
\item Holding $dw_{\alpha}$ fixed, we sample $du_{\alpha}$ and $dv_{\alpha}$
$N_{HS}$ times and for each pair of such values we propagate in time
$\Phi\left(t\right)$ to a new Slater wave-function $\Phi^{\left(k\right)}\left(t+\Delta t\right)=\det\left[\phi_{1}^{\left(k\right)}\left(x_{1}\right)\cdots\phi_{N}^{\left(k\right)}\left(x_{N}\right)\right]$
\begin{equation}
\phi_{n}^{\left(k\right)}\left(t+\Delta t\right)=e^{-\frac{i}{\hbar}\left(\hat{h}dt-i\left(d\hat{h}_{R}+d\hat{h}_{C}+d\hat{h}_{S}\right)\right)}\phi_{n}\left(t\right)
\end{equation}
for $k=1,\dots,N_{HS}$ .
\item Generate from $\Phi^{\left(1\right)},\dots,\Phi^{\left(N_{HS}\right)}$
the one-particle density matrix $\rho_{1}\left(r,r'\right)$ and diagonalize
it:
\begin{equation}
\rho_{1}\left(r,r'\right)=\sum_{n}\tilde{w}_{n}\tilde{\phi}_{n}\left(r\right)\tilde{\phi}_{n}\left(r'\right)^{*},
\end{equation}
(where $\tilde{w}_{1}\ge\tilde{w}_{2}\ge\tilde{w}_{3}\dots$). Now
select the first eigenfunctions $\tilde{\phi}_{n}$ of $\rho_{1}$
and form from them the Slater wave function to be used as the wave
function for the next time step $\Phi\left(t+dt\right)\equiv\det\left[\tilde{\phi}_{1}\cdots\tilde{\phi}_{N}\right]$.
We note that $\Phi\left(t+dt\right)$ is the single determinant wave
function which reproduces the one-body DM $\rho_{1}$as close as possible.
The initial state and the expected values for the Lindblad operator
to be used in the next iteration will thus be be calculated as:
\begin{equation}
\left\langle L^{\alpha}\right\rangle _{t+dt}=\sum_{n=1}^{N}\left\langle \tilde{\phi}_{n}\left|\hat{\ell}^{\alpha}\right|\tilde{\phi}_{n}\right\rangle .
\end{equation}
\end{enumerate}
\end{enumerate}
The last step of the algorithm involves collapsing the Hubbard-Stratonovich
step into a Slater wave function having a similar one-body density
matrix. This step can be generalized and one can retain a wave function
which is a linear combination of $Q\ge1$ determinants that yield
a similar one-body density matrix. In principle, one needs to take
$Q\to\infty$ but in practice we should check that the calculation
is converged with respect to $Q$. In the present paper we do not
attempt to converge the calculation results with respect to $Q$.
In many applications the system is driven to its thermal equal

\section{\label{sec:Applications-to-trapped}Applications to trapped 1D Fermions}

To demonstrate the validity of the method we report calculations on
systems of $N$ non-interacting spin-up Fermions of mass $m=1m_{e}$
(atomic units are used in all reported numerical results) trapped
in a 1D potential $V\left(x\right)$ (see Eq.~(\ref{eq:single-particle-h}))
and using only one Lindblad operator 
\begin{equation}
\hat{\ell}\equiv\sqrt{\frac{m\omega_{\ell}\gamma}{2\hbar N}}\left(\hat{x}+\frac{i}{m\omega_{\ell}}\hat{p}\right)
\end{equation}
to be used in Eq.~(\ref{eq:L-alpha}). Note that $\hat{\ell}$ is
the lowering ladder operator for a harmonic oscillator of frequency
$\omega_{\ell}$(although it is a still also Fermionic operator).
In the results shown below, we use $\omega_{\ell}=1E_{h}/\hbar$ and
$\gamma=0.2E_{h}/\hbar$ and $N=8$ Fermions. The calculation were
carried out using a high-order numerical implementation of the algorithm
depicted in the previous section, where the single particle wave functions
and operators were represented on a Fourier grid and the non-unitary
time propagation was performed using a high-degree interpolating polynomial
in the Newton form.\cite{tal1988high,Kosloff1994} 

\subsection{\label{subsec:A-harmonic-trap}Validation: Fermions in an harmonic
trap}

To demonstrate the validity of the method we apply it to a system
of $N=8$ Fermions having the Hamiltonian of Eq.~(\ref{eq:single-particle-h})
with a Harmonic potential 
\begin{align}
V\left(x\right) & =\frac{1}{2}m\omega^{2}x^{2},\label{eq:vHarmonic}\\
\omega & =\omega_{\ell}=1E_{h}/\hbar.\nonumber 
\end{align}
In this case, the expectation values of the total displacement $X_{t}=\left\langle \sum_{n=1}^{N}\hat{x}_{n}\right\rangle _{t}$
and total momentum $P_{t}=\left\langle \sum_{n=1}^{N}\hat{p}_{n}\right\rangle _{t}$
can be determined analytically directly from the Lindblad equation:
\begin{align}
X_{t}^{an} & =\left(P_{0}\cos\omega t+\frac{P_{0}}{m\omega}\sin\omega t\right)e^{-\frac{\gamma}{2}t}\label{eq:xt}\\
P_{t}^{an} & =\left(P_{0}\cos\omega t-m\omega X_{0}\sin\omega t\right)e^{-\frac{\gamma}{2}t}.\label{eq:pt}
\end{align}
These trajectories are dependent only on the initial values of the
total displacement $X_{0}$ and momentum $P_{0}$ and not explicitly
on the number of electrons $N$ or on other properties of the initial
state. In our demonstration we start from a pure state which is taken
as the a non-stationary Slater wave-function 
\begin{align}
\Phi_{\theta} & =\frac{1}{N!}\det\left[\psi_{1}\left(x_{1}\right)\cdots\psi_{N-1}\left(x_{N-1}\right)\varphi_{\theta}\left(x_{N}\right)\right]\label{eq:InitialPureState}
\end{align}
where $\left\{ \psi_{n}\left(x\right)\right\} _{n=1}^{N+1}$ are the
$N+1$ lowest energy single-particle eigenstates (so-called molecular
orbitals (MO)) of $\hat{h}$, and 
\begin{equation}
\varphi_{\theta}\left(x\right)=\psi_{N}\left(x\right)\cos\theta+\psi_{N+1}\left(x\right)\sin\theta\label{eq:varPhi-Theta}
\end{equation}
is a linear combination involving the highest occupied MO (HOMO) $\psi_{N}\left(x\right)$
and the lowest unoccupied MO (LUMO) $\psi_{N+1}\left(x\right)$. The
angle $\theta$ is taken as $\pi/4$, expressing an equal weight of
these two orbitals. 

In Fig.~\ref{fig:harmonic-XP} we show the analytical trajectory
and the results of 10 independent runs, each based on $N_{traj}=160$
trajectories. The results in the left panel use a time step of $\Delta t=1\hbar/E_{h}$
each employing $N_{HS}=10$ HS iterations while in the right panel
$\Delta t=0.25\hbar/E_{h}$ and $N_{HS}=80$. It can be seen that
the numerical results follow closely the analytical trajectories,
with somewhat improved performance for the smaller time step and more
intensive Hubbard-Stratonovich sampling. The total and kinetic energies
for the trajectories decay to a finite value as $t$ grows. The asymptotic
values for the total and kinetic energies are pushed closer to their
ground state values (which, for this system are $E=32E_{h}$ and $T=16E_{h}$
respectively) as we reduce $\Delta t$ and increase the number of
HS iterations.

\begin{figure}
\includegraphics[width=1\columnwidth]{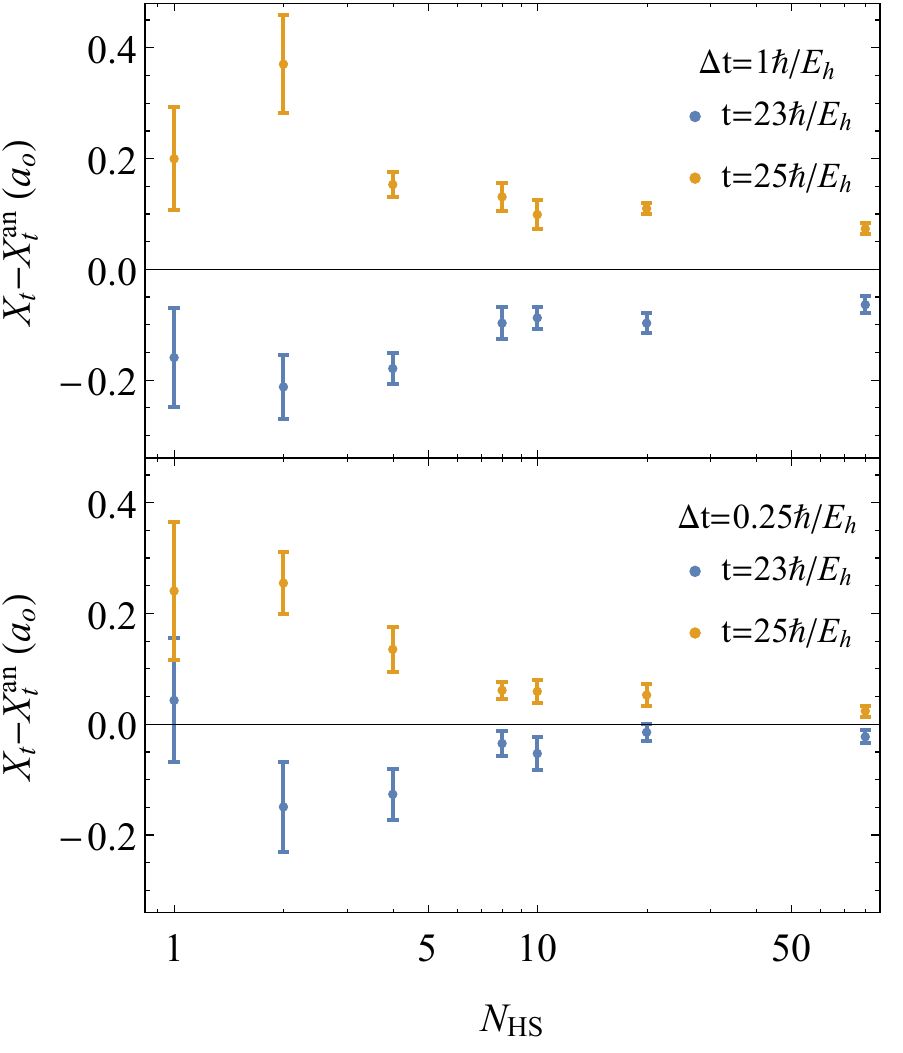}

\caption{\label{fig:Bias}75\% confidence intervals for the difference between
the estimated and analytical displacements at two different values
of $t$ as a function of the number $N_{HS}$ of HS iterations, for
the harmonic system of Section~\ref{subsec:A-harmonic-trap}. Two
time steps are considered: $\Delta t=0.25\hbar/E_{h}$ (top panel)
and $\Delta t=0.125\hbar/E_{h}$ (bottom panel). The confidence intervals
are based on the calculated results from 10 independent runs each
having $N_{traj}=160$ trajectories. }
\end{figure}

A closer look into the accuracy of the dynamics is given in Fig.~\ref{fig:Bias},
where the the 75\% confidence intervals (CIs) for the difference $X_{t}-X_{t}^{an}$
are given at two times, namely $t=23\hbar/E_{h}$ and $t=25\hbar/E_{h}$
as a function of $N_{HS}$ and for two times-steps $\Delta t$. For
$N_{HS}<8$ the results show explicit bias since the error bars of
$N_{HS}\ge8$ are almost non-overlapping with those of $N_{HS}<8$.
For $N_{HS}>8$ the main effect of $N_{HS}$ is reduction of the error
bars (namely improved sampling removes noise). Even for $N_{HS}>8$
the confidence intervals do not include the exact result ($X_{t}-X_{t}^{an}=0$)
showing that a bias exists due to another source, namely the time-step
error. Indeed, as $\Delta t$ decreases from $1$to $0.25$ this bias
decreases by this bias decreases substantially. Hence the time time
step error is the main source of bias for $N_{HS}\ge8$.

Note however that the correct value, namely $X_{t}-X_{t}^{an}=0$,
will not be included in the CI's when we increase $N_{traj}$ due
to the finite-$N_{HS}$ and finite-$\Delta t$ errors which are clearly
noticeable and which can be systematically reduced by increasing $N_{HS}$
and by diminishing $\Delta t$. The results seem converged with respect
to $N_{HS}$ once $N_{HS}>10$ (i.e. although increasing $N_{HS}$
lowers the fluctuation, it does not chan, on the other hand, the main
source of bias is the size of the time step. Once $\Delta t<0.25\hbar/E_{h}$
the 99\% confidence intercal cust the 

\begin{figure*}[t]
\includegraphics[width=1\textwidth]{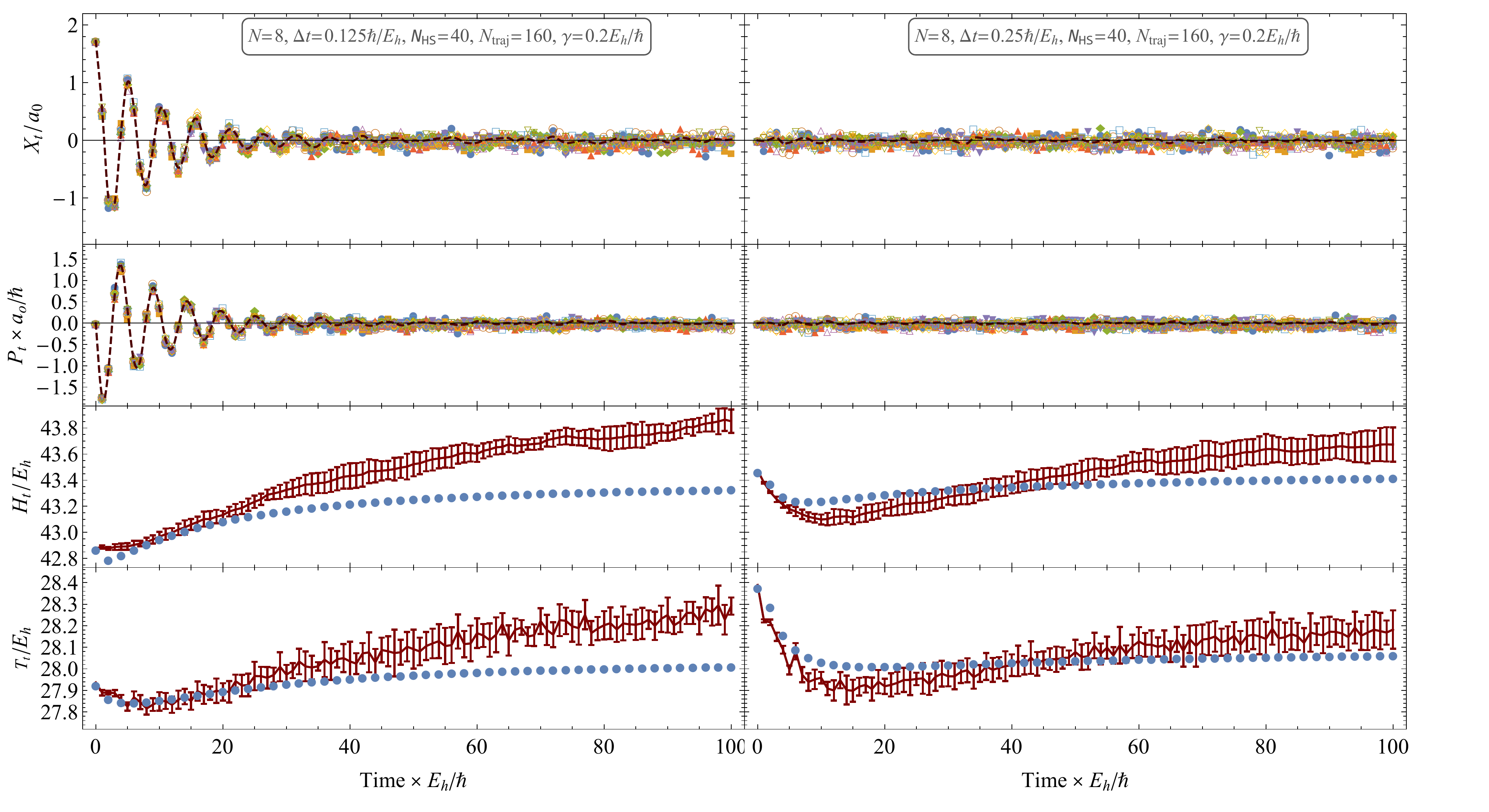}

\caption{\label{fig:dw-XP}The time-dependent total displacement $X_{t}$,
total momentum $P_{t}$, total energy $H_{t}$ and total kinetic energy
$T_{t}$ transients for $8$ non-interacting Fermions in the double-well
trap of Section~\ref{subsec:A-double-well-trap}, starting from a
pure state $\hat{\rho}_{\theta}=\left|\Phi_{\theta}\right\rangle \left\langle \Phi_{\theta}\right|$
with $\theta=\pi/4$ (left panels) and $\theta=\pi/2$ (right panels).
The $X_{t}$ and $P_{t}$ panels show as symbols the calculated results
of 10 independent runs (each having 160 trajectories). The $H_{t}$
and $T_{t}$ results are depicted as statistical error bars centered
on the average over the 10 runs and the blue dots are approximate
transients computed from Eq.~(\ref{eq:SSPopulatio}). }
\end{figure*}

\subsection{\label{subsec:A-double-well-trap}A double-well trap}

As an application of the method, we study $N=8$ Fermions in a double-well
potential obtained by adding to the Harmonic potential of Eq.~(\ref{eq:vHarmonic})
a Gaussian barrier centered at the origin of coordinates: 
\begin{align}
V\left(x\right) & =\frac{1}{2}m\omega^{2}x^{2}+V_{B}e^{-\frac{x^{2}}{2\sigma_{B}^{2}}},\label{eq:vdblWPot}\\
V_{B} & =8E_{h},\,\,\,\sigma_{B}=0.2a_{0}.\nonumber 
\end{align}
In Fig.~\ref{fig:dw-XP} we study the dynamics under similar conditions
of the previous section starting from two different initial pure states
$\hat{\rho}_{\theta}=\left|\Phi_{\theta}\right\rangle \left\langle \Phi_{\theta}\right|$.
On the left panel  the initial state ($\theta=\pi/4$ ) involves a
linear combination of HOMO and LUMO and thus is not an eigenstate
of $\hat{H}$; therefore a damped oscillation in $X$ and $P$ is
observed, which is accompanied by a gradual decrease in the frequency
of oscillation. The energy of the system grows in time, as does the
kinetic energy, indicating that the bath is \emph{injecting} energy
into the system, raising its temperature while at the same time oscillations
are damped due to dephasing. On the right panel we show the transients
corresponding to $\theta=\pi/2$, in which the initial state is an
excited eigenstate of $\hat{H}$ (where the HOMO is replaced by the
LUMO). In an eigenstate there is no motion, so we observe no oscillations
in $X$ and $T$, and it can be supposed that any energy injected
by the bath into the system cannot not stir up observable oscillation
due to the dephasing effects seen in the left panel. The energy here
starts, at early times to decrease but then at $t=11\hbar/E_{h}$
it reverses and starts ascending. The kinetic energy follows this
trend, indicating a tendency for the temperature to initial drop,
reach a minimum somewhat later than the total energy at $t\approx15\hbar/E_{h}$
and then rise at later times. 

We compare these transients to approximate transients based the approximation
that the population of state $i$ is given by: 
\begin{equation}
\dot{n}_{i}\left(t\right)=\sum_{j}\left[\gamma_{ij}n_{j}\left(1-n_{i}\right)-\gamma_{ji}n_{i}\left(1-n_{j}\right)\right],\label{eq:SSPopulatio}
\end{equation}
where $\gamma_{ij}=\sum_{\alpha}\left|\ell_{ij}^{\alpha}\right|^{2}$.\footnote{This equation is obtained by first neglecting the off-diagonal elements
of the DM (expressed as matrix in the eigenstate basis of the Hamiltonian),
which leads to Pauli master equation \cite{schaller2014open} and
then assuming that the populations in states $i$ and $j$ are uncorrelated.} The populations $n_{i}$ enable calculation of the energy and kinetic
energy transients shown as blue dots in Fig.~\ref{fig:dw-XP}. Consider
first the right panel. Here, the initial DM is diagonal so the blue
dots are close to the stochastic calculation, only deviating significantly
when coherences build up at around $t=5\hbar/E_{h}$. While the two
transients are close only at very early times, they both indicate
a non-monotonic behavior of the energy, first cooling and then heating.
For the kinetic energy too there is an agreement at early times where
the system cools at first and then heats up. For the left panel the
initial state is not diagonal so the blue-dot transient transient
breaks off from the more accurate calculation almost immediately.
Again both the accurate and the approximate transients agree qualitatively
that the system is heated by the bath.

\section{\label{sec:Summary}Summary}

In this paper we have introduced a new method for treating the dynamics
of non-interacting Fermions coupled to an external bath. The main
obstacle is the effective inter-particle interactions. We have used
the Hubbard-Stratonovich transformation for reformulation of the unraveled
dynamics to include several types of random walks (three for each
Lindblad operator) which together allow for sampling of the expected
value of the Lindblad operator at a given time. Between different
time-steps a linear combination of $K$ Slater wave functions, reproducing
approximately the the reduced density matrix of the system is formed
(in this work we set $K=1)$. We have shown that this approach allows
for accurate reconstruction of the dynamics of non-interacting Fermions
in a Harmonic oscillator potential well, coupled to a bath through
a specific Lindblad operator. We have also studied the dynamics of
such Fermions in a double-well system, where a non-monotonic behavior
of the energy can be seen when starting from an excited eigenstate
of the Hamiltonian.

This development has the potential of technically enabling a time-dependent
density functional approach for electron dynamics in open systems.
Future development is needed to assess the generality of the results
presented here, implement the option of using a many determinant wave
function (extending the method in this paper where we ``collapse''
to a single determinant state after every time step) and apply the
shifted contour technique for decreasing the Hubbard-Stratonovich
statistical fluctuations.\cite{Rom1997,Baer1998b} Finally, the combination
of the present development with stochastic orbital methods for electronic
structure is an exciting venue.\cite{Baer2013,Baer2013a,Cytter2018,Rabani2015,Neuhauser2017}
\begin{acknowledgments}
This article is submitted to the Festschrift in honor of Prof. Michael
Baer. The second author, Roi Baer, hereby sends his father a happy
80'th birthday with deep love, appreciation and gratitude! Both authors
gratefully thank the Israel Science Foundation Grant No. 189-14 for
kindly funding this research. 
\end{acknowledgments}


\begin{thebibliography}{46}%
\makeatletter
\providecommand \@ifxundefined [1]{%
 \@ifx{#1\undefined}
}%
\providecommand \@ifnum [1]{%
 \ifnum #1\expandafter \@firstoftwo
 \else \expandafter \@secondoftwo
 \fi
}%
\providecommand \@ifx [1]{%
 \ifx #1\expandafter \@firstoftwo
 \else \expandafter \@secondoftwo
 \fi
}%
\providecommand \natexlab [1]{#1}%
\providecommand \enquote  [1]{``#1''}%
\providecommand \bibnamefont  [1]{#1}%
\providecommand \bibfnamefont [1]{#1}%
\providecommand \citenamefont [1]{#1}%
\providecommand \href@noop [0]{\@secondoftwo}%
\providecommand \href [0]{\begingroup \@sanitize@url \@href}%
\providecommand \@href[1]{\@@startlink{#1}\@@href}%
\providecommand \@@href[1]{\endgroup#1\@@endlink}%
\providecommand \@sanitize@url [0]{\catcode `\\12\catcode `\$12\catcode
  `\&12\catcode `\#12\catcode `\^12\catcode `\_12\catcode `\%12\relax}%
\providecommand \@@startlink[1]{}%
\providecommand \@@endlink[0]{}%
\providecommand \url  [0]{\begingroup\@sanitize@url \@url }%
\providecommand \@url [1]{\endgroup\@href {#1}{\urlprefix }}%
\providecommand \urlprefix  [0]{URL }%
\providecommand \Eprint [0]{\href }%
\providecommand \doibase [0]{http://dx.doi.org/}%
\providecommand \selectlanguage [0]{\@gobble}%
\providecommand \bibinfo  [0]{\@secondoftwo}%
\providecommand \bibfield  [0]{\@secondoftwo}%
\providecommand \translation [1]{[#1]}%
\providecommand \BibitemOpen [0]{}%
\providecommand \bibitemStop [0]{}%
\providecommand \bibitemNoStop [0]{.\EOS\space}%
\providecommand \EOS [0]{\spacefactor3000\relax}%
\providecommand \BibitemShut  [1]{\csname bibitem#1\endcsname}%
\let\auto@bib@innerbib\@empty
\bibitem [{\citenamefont {Head-Gordon}\ and\ \citenamefont
  {Tully}(1995)}]{Head-Gordon1995}%
  \BibitemOpen
  \bibfield  {author} {\bibinfo {author} {\bibfnamefont {M.}~\bibnamefont
  {Head-Gordon}}\ and\ \bibinfo {author} {\bibfnamefont {J.~C.}\ \bibnamefont
  {Tully}},\ }\href@noop {} {\bibfield  {journal} {\bibinfo  {journal} {J.
  Chem. Phys.}\ }\textbf {\bibinfo {volume} {103}},\ \bibinfo {pages} {10137}
  (\bibinfo {year} {1995})}\BibitemShut {NoStop}%
\bibitem [{\citenamefont {Schaller}\ \emph {et~al.}(2005)\citenamefont
  {Schaller}, \citenamefont {Pietryga}, \citenamefont {Goupalov}, \citenamefont
  {Petruska}, \citenamefont {Ivanov},\ and\ \citenamefont
  {Klimov}}]{schaller2005breaking}%
  \BibitemOpen
  \bibfield  {author} {\bibinfo {author} {\bibfnamefont {R.~D.}\ \bibnamefont
  {Schaller}}, \bibinfo {author} {\bibfnamefont {J.~M.}\ \bibnamefont
  {Pietryga}}, \bibinfo {author} {\bibfnamefont {S.~V.}\ \bibnamefont
  {Goupalov}}, \bibinfo {author} {\bibfnamefont {M.~A.}\ \bibnamefont
  {Petruska}}, \bibinfo {author} {\bibfnamefont {S.~A.}\ \bibnamefont
  {Ivanov}}, \ and\ \bibinfo {author} {\bibfnamefont {V.~I.}\ \bibnamefont
  {Klimov}},\ }\href@noop {} {\bibfield  {journal} {\bibinfo  {journal}
  {Physical review letters}\ }\textbf {\bibinfo {volume} {95}},\ \bibinfo
  {pages} {196401} (\bibinfo {year} {2005})}\BibitemShut {NoStop}%
\bibitem [{\citenamefont {Baer}\ and\ \citenamefont
  {Billing}(2002)}]{baer2003therole}%
  \BibitemOpen
  \bibinfo {editor} {\bibfnamefont {M.}~\bibnamefont {Baer}}\ and\ \bibinfo
  {editor} {\bibfnamefont {G.~D.}\ \bibnamefont {Billing}},\ eds.,\ \href@noop
  {} {\emph {\bibinfo {title} {The Role of Degenerate States in Chemistry}}},\
  \bibinfo {series} {Advances in Chemical Physics}, Vol.\ \bibinfo {volume}
  {124}\ (\bibinfo  {publisher} {Wiley-Interscience},\ \bibinfo {year}
  {2002})\BibitemShut {NoStop}%
\bibitem [{\citenamefont {Baer}(2006)}]{Baer2006}%
  \BibitemOpen
  \bibfield  {author} {\bibinfo {author} {\bibfnamefont {M.}~\bibnamefont
  {Baer}},\ }\href@noop {} {\emph {\bibinfo {title} {Beyond Born-Oppenheimer:
  electronic non-adiabatic coupling terms and conical intersections}}}\
  (\bibinfo  {publisher} {Wiley},\ \bibinfo {address} {Hoboken, N.J.},\
  \bibinfo {year} {2006})\ pp.\ \bibinfo {pages} {xvii, 234 p.}\BibitemShut
  {Stop}%
\bibitem [{\citenamefont {Gdor}\ \emph {et~al.}(2015)\citenamefont {Gdor},
  \citenamefont {Shapiro}, \citenamefont {Yang}, \citenamefont {Yanover},
  \citenamefont {Lifshitz},\ and\ \citenamefont {Ruhman}}]{Gdor2015}%
  \BibitemOpen
  \bibfield  {author} {\bibinfo {author} {\bibfnamefont {I.}~\bibnamefont
  {Gdor}}, \bibinfo {author} {\bibfnamefont {A.}~\bibnamefont {Shapiro}},
  \bibinfo {author} {\bibfnamefont {C.}~\bibnamefont {Yang}}, \bibinfo {author}
  {\bibfnamefont {D.}~\bibnamefont {Yanover}}, \bibinfo {author} {\bibfnamefont
  {E.}~\bibnamefont {Lifshitz}}, \ and\ \bibinfo {author} {\bibfnamefont
  {S.}~\bibnamefont {Ruhman}},\ }\href@noop {} {\bibfield  {journal} {\bibinfo
  {journal} {ACS Nano}\ }\textbf {\bibinfo {volume} {9}},\ \bibinfo {pages}
  {2138} (\bibinfo {year} {2015})}\BibitemShut {NoStop}%
\bibitem [{\citenamefont {Shenvi}\ and\ \citenamefont
  {Tully}(2012)}]{shenvi2012nonadiabatic}%
  \BibitemOpen
  \bibfield  {author} {\bibinfo {author} {\bibfnamefont {N.}~\bibnamefont
  {Shenvi}}\ and\ \bibinfo {author} {\bibfnamefont {J.~C.}\ \bibnamefont
  {Tully}},\ }\href@noop {} {\bibfield  {journal} {\bibinfo  {journal} {Faraday
  discussions}\ }\textbf {\bibinfo {volume} {157}},\ \bibinfo {pages} {325}
  (\bibinfo {year} {2012})}\BibitemShut {NoStop}%
\bibitem [{\citenamefont {Dong}\ \emph {et~al.}(2015)\citenamefont {Dong},
  \citenamefont {Trivedi}, \citenamefont {Chakrabortty}, \citenamefont
  {Kobayashi}, \citenamefont {Chan}, \citenamefont {Prezhdo},\ and\
  \citenamefont {Loh}}]{dong2015observation}%
  \BibitemOpen
  \bibfield  {author} {\bibinfo {author} {\bibfnamefont {S.}~\bibnamefont
  {Dong}}, \bibinfo {author} {\bibfnamefont {D.}~\bibnamefont {Trivedi}},
  \bibinfo {author} {\bibfnamefont {S.}~\bibnamefont {Chakrabortty}}, \bibinfo
  {author} {\bibfnamefont {T.}~\bibnamefont {Kobayashi}}, \bibinfo {author}
  {\bibfnamefont {Y.}~\bibnamefont {Chan}}, \bibinfo {author} {\bibfnamefont
  {O.~V.}\ \bibnamefont {Prezhdo}}, \ and\ \bibinfo {author} {\bibfnamefont
  {Z.-H.}\ \bibnamefont {Loh}},\ }\href@noop {} {\bibfield  {journal} {\bibinfo
   {journal} {Nano letters}\ }\textbf {\bibinfo {volume} {15}},\ \bibinfo
  {pages} {6875} (\bibinfo {year} {2015})}\BibitemShut {NoStop}%
\bibitem [{\citenamefont {Collini}\ \emph {et~al.}(2010)\citenamefont
  {Collini}, \citenamefont {Wong}, \citenamefont {Wilk}, \citenamefont {Curmi},
  \citenamefont {Brumer},\ and\ \citenamefont
  {Scholes}}]{collini2010coherently}%
  \BibitemOpen
  \bibfield  {author} {\bibinfo {author} {\bibfnamefont {E.}~\bibnamefont
  {Collini}}, \bibinfo {author} {\bibfnamefont {C.~Y.}\ \bibnamefont {Wong}},
  \bibinfo {author} {\bibfnamefont {K.~E.}\ \bibnamefont {Wilk}}, \bibinfo
  {author} {\bibfnamefont {P.~M.}\ \bibnamefont {Curmi}}, \bibinfo {author}
  {\bibfnamefont {P.}~\bibnamefont {Brumer}}, \ and\ \bibinfo {author}
  {\bibfnamefont {G.~D.}\ \bibnamefont {Scholes}},\ }\href@noop {} {\bibfield
  {journal} {\bibinfo  {journal} {Nature}\ }\textbf {\bibinfo {volume} {463}},\
  \bibinfo {pages} {644} (\bibinfo {year} {2010})}\BibitemShut {NoStop}%
\bibitem [{\citenamefont {Romero}\ \emph {et~al.}(2014)\citenamefont {Romero},
  \citenamefont {Augulis}, \citenamefont {Novoderezhkin}, \citenamefont
  {Ferretti}, \citenamefont {Thieme}, \citenamefont {Zigmantas},\ and\
  \citenamefont {Van~Grondelle}}]{romero2014quantum}%
  \BibitemOpen
  \bibfield  {author} {\bibinfo {author} {\bibfnamefont {E.}~\bibnamefont
  {Romero}}, \bibinfo {author} {\bibfnamefont {R.}~\bibnamefont {Augulis}},
  \bibinfo {author} {\bibfnamefont {V.~I.}\ \bibnamefont {Novoderezhkin}},
  \bibinfo {author} {\bibfnamefont {M.}~\bibnamefont {Ferretti}}, \bibinfo
  {author} {\bibfnamefont {J.}~\bibnamefont {Thieme}}, \bibinfo {author}
  {\bibfnamefont {D.}~\bibnamefont {Zigmantas}}, \ and\ \bibinfo {author}
  {\bibfnamefont {R.}~\bibnamefont {Van~Grondelle}},\ }\href@noop {} {\bibfield
   {journal} {\bibinfo  {journal} {Nature physics}\ }\textbf {\bibinfo {volume}
  {10}},\ \bibinfo {pages} {676} (\bibinfo {year} {2014})}\BibitemShut
  {NoStop}%
\bibitem [{\citenamefont {Ingarden}\ \emph {et~al.}(2013)\citenamefont
  {Ingarden}, \citenamefont {Kossakowski},\ and\ \citenamefont
  {Ohya}}]{ingarden2013information}%
  \BibitemOpen
  \bibfield  {author} {\bibinfo {author} {\bibfnamefont {R.~S.}\ \bibnamefont
  {Ingarden}}, \bibinfo {author} {\bibfnamefont {A.}~\bibnamefont
  {Kossakowski}}, \ and\ \bibinfo {author} {\bibfnamefont {M.}~\bibnamefont
  {Ohya}},\ }\href@noop {} {\emph {\bibinfo {title} {Information dynamics and
  open systems: classical and quantum approach}}},\ Vol.~\bibinfo {volume}
  {86}\ (\bibinfo  {publisher} {Springer Science \& Business Media},\ \bibinfo
  {year} {2013})\BibitemShut {NoStop}%
\bibitem [{\citenamefont {Schlosshauer}(2007)}]{schlosshauer2007decoherence}%
  \BibitemOpen
  \bibfield  {author} {\bibinfo {author} {\bibfnamefont {M.~A.}\ \bibnamefont
  {Schlosshauer}},\ }\href@noop {} {\emph {\bibinfo {title} {Decoherence: and
  the quantum-to-classical transition}}}\ (\bibinfo  {publisher} {Springer
  Science \& Business Media},\ \bibinfo {year} {2007})\BibitemShut {NoStop}%
\bibitem [{\citenamefont {Prezhdo}(1999)}]{prezhdo1999mean}%
  \BibitemOpen
  \bibfield  {author} {\bibinfo {author} {\bibfnamefont {O.~V.}\ \bibnamefont
  {Prezhdo}},\ }\href@noop {} {\bibfield  {journal} {\bibinfo  {journal} {J.
  Chem. Phys.}\ }\textbf {\bibinfo {volume} {111}},\ \bibinfo {pages} {8366}
  (\bibinfo {year} {1999})}\BibitemShut {NoStop}%
\bibitem [{\citenamefont {M{\"u}hlbacher}\ and\ \citenamefont
  {Rabani}(2008)}]{muhlbacher2008real}%
  \BibitemOpen
  \bibfield  {author} {\bibinfo {author} {\bibfnamefont {L.}~\bibnamefont
  {M{\"u}hlbacher}}\ and\ \bibinfo {author} {\bibfnamefont {E.}~\bibnamefont
  {Rabani}},\ }\href@noop {} {\bibfield  {journal} {\bibinfo  {journal}
  {Physical review letters}\ }\textbf {\bibinfo {volume} {100}},\ \bibinfo
  {pages} {176403} (\bibinfo {year} {2008})}\BibitemShut {NoStop}%
\bibitem [{\citenamefont {Abramavicius}\ \emph {et~al.}(2009)\citenamefont
  {Abramavicius}, \citenamefont {Palmieri}, \citenamefont {Voronine},
  \citenamefont {Sanda},\ and\ \citenamefont
  {Mukamel}}]{abramavicius2009coherent}%
  \BibitemOpen
  \bibfield  {author} {\bibinfo {author} {\bibfnamefont {D.}~\bibnamefont
  {Abramavicius}}, \bibinfo {author} {\bibfnamefont {B.}~\bibnamefont
  {Palmieri}}, \bibinfo {author} {\bibfnamefont {D.~V.}\ \bibnamefont
  {Voronine}}, \bibinfo {author} {\bibfnamefont {F.}~\bibnamefont {Sanda}}, \
  and\ \bibinfo {author} {\bibfnamefont {S.}~\bibnamefont {Mukamel}},\
  }\href@noop {} {\bibfield  {journal} {\bibinfo  {journal} {Chemical reviews}\
  }\textbf {\bibinfo {volume} {109}},\ \bibinfo {pages} {2350} (\bibinfo {year}
  {2009})}\BibitemShut {NoStop}%
\bibitem [{\citenamefont {Esposito}\ and\ \citenamefont
  {Galperin}(2009)}]{Esposito2009}%
  \BibitemOpen
  \bibfield  {author} {\bibinfo {author} {\bibfnamefont {M.}~\bibnamefont
  {Esposito}}\ and\ \bibinfo {author} {\bibfnamefont {M.}~\bibnamefont
  {Galperin}},\ }\href@noop {} {\bibfield  {journal} {\bibinfo  {journal}
  {Phys. Rev. B}\ }\textbf {\bibinfo {volume} {79}},\ \bibinfo {pages} {205303}
  (\bibinfo {year} {2009})}\BibitemShut {NoStop}%
\bibitem [{\citenamefont {Wilner}\ \emph {et~al.}(2013)\citenamefont {Wilner},
  \citenamefont {Wang}, \citenamefont {Cohen}, \citenamefont {Thoss},\ and\
  \citenamefont {Rabani}}]{wilner2013bistability}%
  \BibitemOpen
  \bibfield  {author} {\bibinfo {author} {\bibfnamefont {E.~Y.}\ \bibnamefont
  {Wilner}}, \bibinfo {author} {\bibfnamefont {H.}~\bibnamefont {Wang}},
  \bibinfo {author} {\bibfnamefont {G.}~\bibnamefont {Cohen}}, \bibinfo
  {author} {\bibfnamefont {M.}~\bibnamefont {Thoss}}, \ and\ \bibinfo {author}
  {\bibfnamefont {E.}~\bibnamefont {Rabani}},\ }\href@noop {} {\bibfield
  {journal} {\bibinfo  {journal} {Physical Review B}\ }\textbf {\bibinfo
  {volume} {88}},\ \bibinfo {pages} {045137} (\bibinfo {year}
  {2013})}\BibitemShut {NoStop}%
\bibitem [{\citenamefont {Cohen}\ \emph {et~al.}(2013)\citenamefont {Cohen},
  \citenamefont {Gull}, \citenamefont {Reichman}, \citenamefont {Millis},\ and\
  \citenamefont {Rabani}}]{cohen2013numerically}%
  \BibitemOpen
  \bibfield  {author} {\bibinfo {author} {\bibfnamefont {G.}~\bibnamefont
  {Cohen}}, \bibinfo {author} {\bibfnamefont {E.}~\bibnamefont {Gull}},
  \bibinfo {author} {\bibfnamefont {D.~R.}\ \bibnamefont {Reichman}}, \bibinfo
  {author} {\bibfnamefont {A.~J.}\ \bibnamefont {Millis}}, \ and\ \bibinfo
  {author} {\bibfnamefont {E.}~\bibnamefont {Rabani}},\ }\href@noop {}
  {\bibfield  {journal} {\bibinfo  {journal} {Physical Review B}\ }\textbf
  {\bibinfo {volume} {87}},\ \bibinfo {pages} {195108} (\bibinfo {year}
  {2013})}\BibitemShut {NoStop}%
\bibitem [{\citenamefont {Schinabeck}\ \emph {et~al.}(2016)\citenamefont
  {Schinabeck}, \citenamefont {Erpenbeck}, \citenamefont {H{\"a}rtle},\ and\
  \citenamefont {Thoss}}]{schinabeck2016hierarchical}%
  \BibitemOpen
  \bibfield  {author} {\bibinfo {author} {\bibfnamefont {C.}~\bibnamefont
  {Schinabeck}}, \bibinfo {author} {\bibfnamefont {A.}~\bibnamefont
  {Erpenbeck}}, \bibinfo {author} {\bibfnamefont {R.}~\bibnamefont
  {H{\"a}rtle}}, \ and\ \bibinfo {author} {\bibfnamefont {M.}~\bibnamefont
  {Thoss}},\ }\href@noop {} {\bibfield  {journal} {\bibinfo  {journal} {Phys.
  Rev. B}\ }\textbf {\bibinfo {volume} {94}},\ \bibinfo {pages} {201407}
  (\bibinfo {year} {2016})}\BibitemShut {NoStop}%
\bibitem [{\citenamefont {Runge}\ and\ \citenamefont
  {Gross}(1984)}]{Runge1984}%
  \BibitemOpen
  \bibfield  {author} {\bibinfo {author} {\bibfnamefont {E.}~\bibnamefont
  {Runge}}\ and\ \bibinfo {author} {\bibfnamefont {E.~K.~U.}\ \bibnamefont
  {Gross}},\ }\href@noop {} {\bibfield  {journal} {\bibinfo  {journal} {Phys.
  Rev. Lett.}\ }\textbf {\bibinfo {volume} {52}},\ \bibinfo {pages} {997}
  (\bibinfo {year} {1984})}\BibitemShut {NoStop}%
\bibitem [{\citenamefont {Burke}\ \emph {et~al.}(2005)\citenamefont {Burke},
  \citenamefont {Car},\ and\ \citenamefont {Gebauer}}]{Burke2005}%
  \BibitemOpen
  \bibfield  {author} {\bibinfo {author} {\bibfnamefont {K.}~\bibnamefont
  {Burke}}, \bibinfo {author} {\bibfnamefont {R.}~\bibnamefont {Car}}, \ and\
  \bibinfo {author} {\bibfnamefont {R.}~\bibnamefont {Gebauer}},\ }\href@noop
  {} {\bibfield  {journal} {\bibinfo  {journal} {Phys. Rev. Lett.}\ }\textbf
  {\bibinfo {volume} {94}},\ \bibinfo {pages} {146805} (\bibinfo {year}
  {2005})}\BibitemShut {NoStop}%
\bibitem [{\citenamefont {Kurth}\ \emph {et~al.}(2005)\citenamefont {Kurth},
  \citenamefont {Stefanucci}, \citenamefont {Almbladh}, \citenamefont {Rubio},\
  and\ \citenamefont {Gross}}]{Kurth2005}%
  \BibitemOpen
  \bibfield  {author} {\bibinfo {author} {\bibfnamefont {S.}~\bibnamefont
  {Kurth}}, \bibinfo {author} {\bibfnamefont {G.}~\bibnamefont {Stefanucci}},
  \bibinfo {author} {\bibfnamefont {C.~O.}\ \bibnamefont {Almbladh}}, \bibinfo
  {author} {\bibfnamefont {A.}~\bibnamefont {Rubio}}, \ and\ \bibinfo {author}
  {\bibfnamefont {E.~K.~U.}\ \bibnamefont {Gross}},\ }\href@noop {} {\bibfield
  {journal} {\bibinfo  {journal} {Phys. Rev. B}\ }\textbf {\bibinfo {volume}
  {72}},\ \bibinfo {pages} {035308} (\bibinfo {year} {2005})}\BibitemShut
  {NoStop}%
\bibitem [{\citenamefont {Zheng}\ \emph {et~al.}(2007)\citenamefont {Zheng},
  \citenamefont {Zhao},\ and\ \citenamefont {Truhlar}}]{Zheng2007}%
  \BibitemOpen
  \bibfield  {author} {\bibinfo {author} {\bibfnamefont {J.~J.}\ \bibnamefont
  {Zheng}}, \bibinfo {author} {\bibfnamefont {Y.}~\bibnamefont {Zhao}}, \ and\
  \bibinfo {author} {\bibfnamefont {D.~G.}\ \bibnamefont {Truhlar}},\
  }\href@noop {} {\bibfield  {journal} {\bibinfo  {journal} {J. Chem. Theory
  Comput.}\ }\textbf {\bibinfo {volume} {3}},\ \bibinfo {pages} {569} (\bibinfo
  {year} {2007})}\BibitemShut {NoStop}%
\bibitem [{\citenamefont {Pershin}\ \emph {et~al.}(2008)\citenamefont
  {Pershin}, \citenamefont {Dubi},\ and\ \citenamefont
  {Di~Ventra}}]{Pershin2008}%
  \BibitemOpen
  \bibfield  {author} {\bibinfo {author} {\bibfnamefont {Y.~V.}\ \bibnamefont
  {Pershin}}, \bibinfo {author} {\bibfnamefont {Y.}~\bibnamefont {Dubi}}, \
  and\ \bibinfo {author} {\bibfnamefont {M.}~\bibnamefont {Di~Ventra}},\
  }\href@noop {} {\bibfield  {journal} {\bibinfo  {journal} {Phys. Rev. B}\
  }\textbf {\bibinfo {volume} {78}},\  (\bibinfo {year} {2008})}\BibitemShut
  {NoStop}%
\bibitem [{\citenamefont {Yuen-Zhou}\ \emph {et~al.}(2010)\citenamefont
  {Yuen-Zhou}, \citenamefont {Tempel}, \citenamefont {Rodr{\'\i}guez-Rosario},\
  and\ \citenamefont {Aspuru-Guzik}}]{Yuen-Zhou2010}%
  \BibitemOpen
  \bibfield  {author} {\bibinfo {author} {\bibfnamefont {J.}~\bibnamefont
  {Yuen-Zhou}}, \bibinfo {author} {\bibfnamefont {D.~G.}\ \bibnamefont
  {Tempel}}, \bibinfo {author} {\bibfnamefont {C.~A.}\ \bibnamefont
  {Rodr{\'\i}guez-Rosario}}, \ and\ \bibinfo {author} {\bibfnamefont
  {A.}~\bibnamefont {Aspuru-Guzik}},\ }\href@noop {} {\bibfield  {journal}
  {\bibinfo  {journal} {Phys. Rev. Lett.}\ }\textbf {\bibinfo {volume} {104}},\
  \bibinfo {pages} {043001} (\bibinfo {year} {2010})}\BibitemShut {NoStop}%
\bibitem [{Note1()}]{Note1}%
  \BibitemOpen
  \bibinfo {note} {This is true when the electrons interact with the bath
  through the one-body density matrix, which the case of interest here and in
  most practical applications. There exists an important class of problems in
  which the electron interaction with the bath is ``linear'' with the particle
  creation/destruction operators allowing an easier TDDFT adaption (see \cite
  {Kurth2005}).}\BibitemShut {Stop}%
\bibitem [{\citenamefont {Lindblad}(1976)}]{Lindblad1976}%
  \BibitemOpen
  \bibfield  {author} {\bibinfo {author} {\bibfnamefont {G.}~\bibnamefont
  {Lindblad}},\ }\href@noop {} {\bibfield  {journal} {\bibinfo  {journal}
  {Commun. Math. Phys.}\ }\textbf {\bibinfo {volume} {48}},\ \bibinfo {pages}
  {119} (\bibinfo {year} {1976})}\BibitemShut {NoStop}%
\bibitem [{\citenamefont {Gorini}\ \emph {et~al.}(1976)\citenamefont {Gorini},
  \citenamefont {Kossakowski},\ and\ \citenamefont
  {Sudarshan}}]{gorini1976completely}%
  \BibitemOpen
  \bibfield  {author} {\bibinfo {author} {\bibfnamefont {V.}~\bibnamefont
  {Gorini}}, \bibinfo {author} {\bibfnamefont {A.}~\bibnamefont {Kossakowski}},
  \ and\ \bibinfo {author} {\bibfnamefont {E.~C.~G.}\ \bibnamefont
  {Sudarshan}},\ }\href@noop {} {\bibfield  {journal} {\bibinfo  {journal}
  {Journal of Mathematical Physics}\ }\textbf {\bibinfo {volume} {17}},\
  \bibinfo {pages} {821} (\bibinfo {year} {1976})}\BibitemShut {NoStop}%
\bibitem [{\citenamefont {Breuer}\ and\ \citenamefont
  {Petruccione}(2002)}]{Breuer2002}%
  \BibitemOpen
  \bibfield  {author} {\bibinfo {author} {\bibfnamefont {H.-P.}\ \bibnamefont
  {Breuer}}\ and\ \bibinfo {author} {\bibfnamefont {F.}~\bibnamefont
  {Petruccione}},\ }\href@noop {} {\emph {\bibinfo {title} {The theory of open
  quantum systems}}}\ (\bibinfo  {publisher} {Oxford University Press},\
  \bibinfo {address} {Oxford ; New York},\ \bibinfo {year} {2002})\ pp.\
  \bibinfo {pages} {xxi, 625 p.}\BibitemShut {Stop}%
\bibitem [{\citenamefont {Schaller}(2014)}]{schaller2014open}%
  \BibitemOpen
  \bibfield  {author} {\bibinfo {author} {\bibfnamefont {G.}~\bibnamefont
  {Schaller}},\ }\href@noop {} {\emph {\bibinfo {title} {Open quantum systems
  far from equilibrium}}},\ Vol.\ \bibinfo {volume} {881}\ (\bibinfo
  {publisher} {Springer},\ \bibinfo {year} {2014})\BibitemShut {NoStop}%
\bibitem [{\citenamefont {Alicki}\ and\ \citenamefont
  {Lendi}(2007)}]{Alicki2007}%
  \BibitemOpen
  \bibfield  {author} {\bibinfo {author} {\bibfnamefont {R.}~\bibnamefont
  {Alicki}}\ and\ \bibinfo {author} {\bibfnamefont {K.}~\bibnamefont {Lendi}},\
  }\href@noop {} {\emph {\bibinfo {title} {Quantum Dynamical Semigroups and
  Applications}}},\ \bibinfo {series} {Lecture notes in Physics}, Vol.\
  \bibinfo {volume} {717}\ (\bibinfo  {publisher} {Springer, Berlin
  Heidelberg},\ \bibinfo {year} {2007})\ pp.\ \bibinfo {pages}
  {1--94}\BibitemShut {NoStop}%
\bibitem [{\citenamefont {Carmichael}(1993)}]{carmichael1993quantum}%
  \BibitemOpen
  \bibfield  {author} {\bibinfo {author} {\bibfnamefont {H.}~\bibnamefont
  {Carmichael}},\ }\href@noop {} {\bibfield  {journal} {\bibinfo  {journal}
  {Physical review letters}\ }\textbf {\bibinfo {volume} {70}},\ \bibinfo
  {pages} {2273} (\bibinfo {year} {1993})}\BibitemShut {NoStop}%
\bibitem [{\citenamefont {Wiseman}\ and\ \citenamefont
  {Milburn}(1993)}]{wiseman1993quantum}%
  \BibitemOpen
  \bibfield  {author} {\bibinfo {author} {\bibfnamefont {H.~M.}\ \bibnamefont
  {Wiseman}}\ and\ \bibinfo {author} {\bibfnamefont {G.~J.}\ \bibnamefont
  {Milburn}},\ }\href@noop {} {\bibfield  {journal} {\bibinfo  {journal} {Phys.
  Rev. A}\ }\textbf {\bibinfo {volume} {47}},\ \bibinfo {pages} {642} (\bibinfo
  {year} {1993})}\BibitemShut {NoStop}%
\bibitem [{\citenamefont {Dalibard}\ \emph {et~al.}(1992)\citenamefont
  {Dalibard}, \citenamefont {Castin},\ and\ \citenamefont
  {M{\o}lmer}}]{dalibard1992wave}%
  \BibitemOpen
  \bibfield  {author} {\bibinfo {author} {\bibfnamefont {J.}~\bibnamefont
  {Dalibard}}, \bibinfo {author} {\bibfnamefont {Y.}~\bibnamefont {Castin}}, \
  and\ \bibinfo {author} {\bibfnamefont {K.}~\bibnamefont {M{\o}lmer}},\
  }\href@noop {} {\bibfield  {journal} {\bibinfo  {journal} {Physical review
  letters}\ }\textbf {\bibinfo {volume} {68}},\ \bibinfo {pages} {580}
  (\bibinfo {year} {1992})}\BibitemShut {NoStop}%
\bibitem [{\citenamefont {Gisin}\ and\ \citenamefont
  {Percival}(1992)}]{Gisin1992}%
  \BibitemOpen
  \bibfield  {author} {\bibinfo {author} {\bibfnamefont {N.}~\bibnamefont
  {Gisin}}\ and\ \bibinfo {author} {\bibfnamefont {I.~C.}\ \bibnamefont
  {Percival}},\ }\href@noop {} {\bibfield  {journal} {\bibinfo  {journal}
  {Journal of Physics a-Mathematical and General}\ }\textbf {\bibinfo {volume}
  {25}},\ \bibinfo {pages} {5677} (\bibinfo {year} {1992})}\BibitemShut
  {NoStop}%
\bibitem [{\citenamefont {Stratonovich}(1957)}]{stratonovich1957rl}%
  \BibitemOpen
  \bibfield  {author} {\bibinfo {author} {\bibfnamefont {R.}~\bibnamefont
  {Stratonovich}},\ }\href@noop {} {\bibfield  {journal} {\bibinfo  {journal}
  {Dokl. Akad. Nauk SSSR}\ }\textbf {\bibinfo {volume} {115}},\ \bibinfo
  {pages} {1097} (\bibinfo {year} {1957})}\BibitemShut {NoStop}%
\bibitem [{\citenamefont {Hubbard}(1959)}]{Hubbard1959}%
  \BibitemOpen
  \bibfield  {author} {\bibinfo {author} {\bibfnamefont {J.}~\bibnamefont
  {Hubbard}},\ }\href@noop {} {\bibfield  {journal} {\bibinfo  {journal} {Phys.
  Rev. Lett.}\ }\textbf {\bibinfo {volume} {3}},\ \bibinfo {pages} {77}
  (\bibinfo {year} {1959})}\BibitemShut {NoStop}%
\bibitem [{\citenamefont {Tal-Ezer}(1988)}]{tal1988high}%
  \BibitemOpen
  \bibfield  {author} {\bibinfo {author} {\bibfnamefont {H.}~\bibnamefont
  {Tal-Ezer}},\ }\href@noop {} {\emph {\bibinfo {title} {High degree
  interpolation polynomial in Newton form}}},\ \bibinfo {type} {Tech. Rep.}\
  \bibinfo {number} {88-39}\ (\bibinfo  {institution} {NASA (ICASE Report)},\
  \bibinfo {year} {1988})\BibitemShut {NoStop}%
\bibitem [{\citenamefont {Kosloff}(1994)}]{Kosloff1994}%
  \BibitemOpen
  \bibfield  {author} {\bibinfo {author} {\bibfnamefont {R.}~\bibnamefont
  {Kosloff}},\ }\href@noop {} {\bibfield  {journal} {\bibinfo  {journal} {Annu.
  Rev. Phys. Chem.}\ }\textbf {\bibinfo {volume} {45}},\ \bibinfo {pages} {145}
  (\bibinfo {year} {1994})}\BibitemShut {NoStop}%
\bibitem [{Note2()}]{Note2}%
  \BibitemOpen
  \bibinfo {note} {This equation is obtained by first neglecting the
  off-diagonal elements of the DM (expressed as matrix in the eigenstate basis
  of the Hamiltonian), which leads to Pauli master equation \cite
  {schaller2014open} and then assuming that the populations in states $i$ and
  $j$ are uncorrelated.}\BibitemShut {Stop}%
\bibitem [{\citenamefont {Rom}\ \emph {et~al.}(1997)\citenamefont {Rom},
  \citenamefont {Charutz},\ and\ \citenamefont {Neuhauser}}]{Rom1997}%
  \BibitemOpen
  \bibfield  {author} {\bibinfo {author} {\bibfnamefont {N.}~\bibnamefont
  {Rom}}, \bibinfo {author} {\bibfnamefont {D.~M.}\ \bibnamefont {Charutz}}, \
  and\ \bibinfo {author} {\bibfnamefont {D.}~\bibnamefont {Neuhauser}},\
  }\href@noop {} {\bibfield  {journal} {\bibinfo  {journal} {Chem. Phys.
  Lett.}\ }\textbf {\bibinfo {volume} {270}},\ \bibinfo {pages} {382} (\bibinfo
  {year} {1997})}\BibitemShut {NoStop}%
\bibitem [{\citenamefont {Baer}\ \emph {et~al.}(1998)\citenamefont {Baer},
  \citenamefont {Head-Gordon},\ and\ \citenamefont {Neuhauser}}]{Baer1998b}%
  \BibitemOpen
  \bibfield  {author} {\bibinfo {author} {\bibfnamefont {R.}~\bibnamefont
  {Baer}}, \bibinfo {author} {\bibfnamefont {M.~P.}\ \bibnamefont
  {Head-Gordon}}, \ and\ \bibinfo {author} {\bibfnamefont {D.}~\bibnamefont
  {Neuhauser}},\ }\href@noop {} {\bibfield  {journal} {\bibinfo  {journal} {J.
  Chem. Phys.}\ }\textbf {\bibinfo {volume} {109}},\ \bibinfo {pages} {6219}
  (\bibinfo {year} {1998})},\ \bibinfo {note} {rBaer-Publication}\BibitemShut
  {NoStop}%
\bibitem [{\citenamefont {Baer}\ \emph {et~al.}(2013)\citenamefont {Baer},
  \citenamefont {Neuhauser},\ and\ \citenamefont {Rabani}}]{Baer2013}%
  \BibitemOpen
  \bibfield  {author} {\bibinfo {author} {\bibfnamefont {R.}~\bibnamefont
  {Baer}}, \bibinfo {author} {\bibfnamefont {D.}~\bibnamefont {Neuhauser}}, \
  and\ \bibinfo {author} {\bibfnamefont {E.}~\bibnamefont {Rabani}},\ }\href
  {\doibase 10.1103/PhysRevLett.111.106402} {\bibfield  {journal} {\bibinfo
  {journal} {Phys. Rev. Lett.}\ }\textbf {\bibinfo {volume} {111}},\ \bibinfo
  {pages} {106402} (\bibinfo {year} {2013})},\ \bibinfo {note}
  {rBaer-Publication}\BibitemShut {NoStop}%
\bibitem [{\citenamefont {Baer}\ and\ \citenamefont
  {Rabani}(2013)}]{Baer2013a}%
  \BibitemOpen
  \bibfield  {author} {\bibinfo {author} {\bibfnamefont {R.}~\bibnamefont
  {Baer}}\ and\ \bibinfo {author} {\bibfnamefont {E.}~\bibnamefont {Rabani}},\
  }\href@noop {} {\bibfield  {journal} {\bibinfo  {journal} {J. Chem. Phys.}\
  }\textbf {\bibinfo {volume} {138}},\ \bibinfo {pages} {051102} (\bibinfo
  {year} {2013})},\ \bibinfo {note} {rBaer-Publication}\BibitemShut {NoStop}%
\bibitem [{\citenamefont {Cytter}\ \emph {et~al.}(2018)\citenamefont {Cytter},
  \citenamefont {Rabani}, \citenamefont {Neuhauser},\ and\ \citenamefont
  {Baer}}]{Cytter2018}%
  \BibitemOpen
  \bibfield  {author} {\bibinfo {author} {\bibfnamefont {Y.}~\bibnamefont
  {Cytter}}, \bibinfo {author} {\bibfnamefont {E.}~\bibnamefont {Rabani}},
  \bibinfo {author} {\bibfnamefont {D.}~\bibnamefont {Neuhauser}}, \ and\
  \bibinfo {author} {\bibfnamefont {R.}~\bibnamefont {Baer}},\ }\href@noop {}
  {\bibfield  {journal} {\bibinfo  {journal} {arXiv:1801.02163
  [cond-mat.mtrl-sci]}\ } (\bibinfo {year} {2018})},\ \bibinfo {note}
  {rBaer-Publication}\BibitemShut {NoStop}%
\bibitem [{\citenamefont {Rabani}\ \emph {et~al.}(2015)\citenamefont {Rabani},
  \citenamefont {Baer},\ and\ \citenamefont {Neuhauser}}]{Rabani2015}%
  \BibitemOpen
  \bibfield  {author} {\bibinfo {author} {\bibfnamefont {E.}~\bibnamefont
  {Rabani}}, \bibinfo {author} {\bibfnamefont {R.}~\bibnamefont {Baer}}, \ and\
  \bibinfo {author} {\bibfnamefont {D.}~\bibnamefont {Neuhauser}},\ }\href@noop
  {} {\bibfield  {journal} {\bibinfo  {journal} {Phys. Rev. B}\ }\textbf
  {\bibinfo {volume} {91}},\ \bibinfo {pages} {235302} (\bibinfo {year}
  {2015})},\ \bibinfo {note} {rBaer-Publication}\BibitemShut {NoStop}%
\bibitem [{\citenamefont {Neuhauser}\ \emph {et~al.}(2017)\citenamefont
  {Neuhauser}, \citenamefont {Baer},\ and\ \citenamefont
  {Zgid}}]{Neuhauser2017}%
  \BibitemOpen
  \bibfield  {author} {\bibinfo {author} {\bibfnamefont {D.}~\bibnamefont
  {Neuhauser}}, \bibinfo {author} {\bibfnamefont {R.}~\bibnamefont {Baer}}, \
  and\ \bibinfo {author} {\bibfnamefont {D.}~\bibnamefont {Zgid}},\ }\href@noop
  {} {\bibfield  {journal} {\bibinfo  {journal} {J. Chem. Theory Comput.}\
  }\textbf {\bibinfo {volume} {13}},\ \bibinfo {pages} {5396} (\bibinfo {year}
  {2017})},\ \bibinfo {note} {rBaer-Publication}\BibitemShut {NoStop}%
\end{thebibliography}
%

\end{document}